\documentstyle[12pt]{JHEP3}

\title{ Generalised Penrose Limits and PP-Waves
 } 
\author{Harvendra Singh\footnote{singh@mri.ernet.in} \\
Harish-Chandra Research Institute,\\
Chhatnag Rd., Jhunsi, Allahabad-211 019, India 
\footnote{Current address: Department of Physics, 
Indian Institute of Technology, Guwahati-781 039 (India) 
[Email: hsingh@iitg.ernet.in]} }
\abstract{ In this paper
we construct generalised Penrose limits for the solutions  of massive 
type IIA supergravity. We consider the Freund-Rubin type solution
and apply these {\it massive} Penrose limits and obtain supersymmetric
 pp-wave  which is  standard type IIA  background.
 The results in this paper can be easily generalised for the cases of
gauged supergravities. }
\begin{document}
\def\be{\begin{equation}} \def\ee{\end{equation}}
\def\bea{\begin{eqnarray}} \def\eea{\end{eqnarray}} \def\ba{\begin{array}}
\def\ea{\end{array}} \def\ben{\begin{enumerate}} \def\een{\end{enumerate}}
\def\nab{\bigtriangledown} \def\tpi{\tilde\Phi} \def\nnu{\nonumber}
\def\lll{\label}
\newcommand{\eqn}[1]{(\ref{#1})}
\def\cC{{\cal C}} 
\def\cG{{\cal G}} 
\def\cd{{\cal D}} 
\def\a{\alpha}
\def\b{\beta}
\def\g{\gamma}\def\G{\Gamma}
\def\d{\delta}\def\D{\Delta}
\def\ep{\epsilon}
\def\e{\eta}
\def\z{\zeta}
\def\t{\theta}\def\T{\Theta}
\def\l{\lambda}\def\L{\Lambda}
\def\m{\mu}
\def\f{\phi}\def\F{\Phi}
\def\n{\nu}
\def\p{\psi}\def\P{\Psi}
\def\r{\rho}
\def\s{\sigma}\def\S{\Sigma}
\def\ta{\tau}
\def\x{\chi}
\def\o{\omega}\def\O{\Omega}
\def\k{\kappa}
\def\pa {\partial}
\def\ov{\over}
\def\br{\nonumber\\}
\def\ud{\underline}
\section{Introduction}
The maximally supersymmetric plane-fronted parallel (pp) wave
configurations in string theory 
can generically be obtained by applying the Penrose 
limits \cite{penrose,gueven,blau,blau1} \footnote{More precisely the 
Penrose limits of Blau 
et. al. \cite{blau,blau1} where we zoom in onto the null geodesics 
involving a direction along the sphere. If the null geodesic is such 
that it does not involve the sphere then we obtain plane-waves after 
taking the Penrose limit.} on the solutions of the type $AdS_p\times S^q$. 
Although,  unlike the flat spacetime  these supersymmetric
pp-waves are (asymptotically) non-flat geometries, 
nevertheless string theory in these
backgrounds becomes exactly solvable theory in 
lightcone gauge\cite{matsaev,matsaev1}.
Also the pp-wave spacetime in bulk has useful consequences for
dual conformal field theories on boundary \cite{maldacena}. 
Under BMN correspondence \cite{bmn} the operators in the CFT with 
large $U(1)$ R-charge, $J$, are dual to type IIB closed string excitations 
in a pp-wave background
spacetime.\footnote{ 
Several useful works have already been carried out in the  
references 
\cite{mukhi,gomis,zayas,billo,kim,takaya,das,chu,clp,beren,dabhol,
hull,lee,hatsuda,kumar, singhpp, alish,alok,spal,rnayak, sanjay1,sanjay2,
Chandrasekhar:2003fq, Stanciu:2003bn, stefanski,ohta}.} 
The supersymmetric 
pp-wave vacua have  been obtained in lower dimensional supergravities 
as well \cite{mees,mees1}. In general, there is an enhancement of the 
supersymmetries after taking Penrose limits on the anti-de
Sitter  vacua.

Unlike standard supergravities, the massive or gauged
supergravities cannot admit pp-wave vacua
due to the presence of cosmological constant (or a potential). 
Several massive and gauged
supergravities admit  $AdS \times S$ type solutions, yet the
Penrose limits for them are unexplored.
The Penrose limits 
\cite{gueven, blau1} map anti-de Sitter solutions of a supergravity
theory to 
the pp-wave solutions of the same supergravity, generically accompanied with 
the enhancement of
supersymmetry. The problem with massive/gauged supergravities is that 
they admit $AdS\times S$ vacua but cannot 
have pp-wave solutions and hence there cannot be  pp-wave limits which 
interpolate between them. Our aim in this work is to
understand the map between anti-de Sitter solutions
  of massive/gauged supergravities 
and the pp-waves under some {\it generalised} Penrose limits. 

We choose an explicit example of massive type IIA supergravity \cite{roma}
in ten dimensions which fits  well to exploit our proposal which is 
the following. First we make two observations about this theory;

i) {\it The
theory admits $AdS\times S$ solutions, although they are not
supersymmetric. }

ii) {\it The  action of this supergravity theory has a scaling 
symmetry under which mass
parameter $m$ also gets scaled.} 

We exploit this scaling to generalise
the Penrose limits \cite{blau,blau1} such that when limits are applied on
the $AdS\times S$ solutions of a gauged supergravity we get pp-wave 
solutions of the standard 
(ungauged) supergravity.

The paper is organised as follows. We discuss the Penrose
limits for massive type IIA supergravity in section-2. 
 We consider non-supersymmetric 
anti-de Sitter solution of massive type IIA
supergravity and apply {\it massive} Penrose limits to obtain 
pp-wave background.
In this way we obtain $1/2$  supersymmetric pp-wave of type IIA
supergravity.  We also remark on the corresponding  Matrix model.  
In section-3 we discuss some cases of gauged supergravities
where such an exercise could be carried out. 
 The last section
section-4 is for the conclusions.

\section{A massive Penrose limit}
The Penrose limits \cite{blau1,penrose,gueven} of 
anti-de Sitter spacetimes $
AdS_p\times S^q $ along any null geodesic with a
generic orbit (i.e. with a non-zero component along sphere) leads to
a pp-wave geometry. The procedure is well described in
\cite{blau1} where maximally supersymmetric pp-wave solutions are
obtained. Our interest here is to study Penrose limits for the
solutions of massive type IIA supergravity for which there exist
anti-de Sitter solutions but has no supersymmetry \cite{roma}. This
theory has an explicit cosmological constant which is proportional to
the square of the mass parameter $m$.  In this supergravity we could 
define a scaling limit on the supergravity fields as,
\bea\label{pen}
 && g_{\m\n}\to \xi^{-2} g_{\m\n} \br
&& \f\to\f\br
&& A_{(p)} \to \xi^{-p} A_{(p)} 
\eea
accompanied with the scaling
\be
 m\to \xi~ m
\ee
 under which the action scales homogeneously, see Appendix. The
 scale parameter  $\xi$ has to be  strictly positive. This 
scaling symmetry 
could be used to tune the mass parameter $m$ in the theory. The 
 generalised or {\it massive } Penrose limits could then be described as 
\bea\lll{pen1}
 && \bar g_{\m\n}=\lim\limits_{\O\to0} \O^{-2} g_{\m\n}(\O) \br
&&\bar\f =\lim\limits_{\O\to0} \f(\O)\br
&& \bar A_{(p)} =\lim\limits_{\O\to0} \O^{-p} A_{(p)}(\O) \br
&&\bar m=\lim\limits_{\O\to0} \O~ m
\eea
where parameter $\O$ is strictly positive. It can be easily noted 
that such a pp-wave limit takes 
mass parameter to zero value and hence will give us pp-wave solutions of 
ordinary type 
IIA string theory starting from  $AdS\times S$ solutions of massive
type IIA supergravity. Note that the limits \eqn{pen1} are well defined 
when the massive IIA supergravity action is written in such a form that 
massless limit could be taken. Thus if there is a solution with the local 
data 
$(AdS, g, \f, A_p; m)$ by implementing the limits \eqn{pen1} we get to 
new data
$(W, \bar g, \bar\f,\bar A_p;\bar m=0)$ which is a pp-wave 
solution of
the ordinary type IIA string
theory.  
 In the next secion we will explicitly cover this aspect by taking 
an example of massive type IIA supergravity solution. These massive Penrose limits 
can be generalised in a straightforward manner
for the case of gauged supergravities as well.  
It should however be noted that the limits \eqn{pen1} are the same as in 
the case of Blau et al \cite{blau, blau1} except in the last equation 
which scales the mass parameter of the massive or gauged supergravity. 
 
\subsection{The Freund-Rubin solution}

The massive type IIA supergravity, action \eqn{massive2a}, has no maximally 
supersymmetric
flat vacua. Due to the cosmological constant this theory
 cannot have pp-wave solutions either.
However, it admits half-supersymmetric D8-brane solutions
\cite{polwit} and D8 bound states with $B$-field \cite{singhd8,singhd6d8}. 
The field equations do also have non-supersymmetric Freund-Rubin type solution
$AdS_4\times S^6$ \cite{roma}
describing the compactification to four dimensions. This solution is given
by
\bea\lll{ss21}
&&g:= l_1^2 \big[ -dt^2 + sin^2t \left({dr^2\over 1+r^2} +r^2
  d\O_2^2\right)\big] + l_2^2\big[ d\psi^2+sin^2\p d\O_5^2\big] \br
&& F_{(4)}=\sqrt{5}~ m {\rm Vol}(AdS_4(l_1)),~~~ \f=0 
 \eea
where $m$ is the mass parameter,
\be\lll{2a2}
l_1={\sqrt{2}/m},~~~~~  l_2={\sqrt{5}/m }\ee
are the radii of the
$AdS_4$ and $S^6$ respectively, and $d\O_n^2$ is the unit $n$-sphere
metric. The volume 
$$ {\rm Vol}(AdS_4(l_1)) = l_1^4 sin^3t{r^2\ov\sqrt{1+r^2}} dtdr\o_2$$
with $\o_2$
being the volume element of a unit 2-sphere.
The rest of the background fields are vanishing in the above. 
Now we would like to change coordinates in the $(\p,t)$ plane to
\be
U=\p + \rho t ,~~~~V=\p - \r t,
\ee
in terms of which the background becomes,
  \bea\lll{s21}
&&l_2^{-2} g:= dUdV + \rho^2 sin^2\left({U-V\over2\r}\right)
  \left({dr^2\over 
1+r^2} +r^2  d\O_2^2\right) + sin^2\left(U+V\ov2\right) d\O_5^2 \br
&& l_2^{-3} F_{(4)}={2\ov\r~l_1} Vol({AdS_4(U,V)}),~~~ \f=0 
 \eea
where $\r=l_1/l_2=\sqrt{2/5}$ is the ratio of the radii of curvatures.
 We now rescale the coordinates as 
\bea
U=u,~~~V={v\ov (l_2)^2},~~~ Y^a={y^a\ov l_2}
\eea
and take the Penrose limits \eqn{pen1},  $l_2\to\infty$ i.e. 
large radius limit,
 along the null geodesic
parametrised by $U$.\footnote{ Since the curvature of spacetime is 
coupled to the mass parameter $m$, see \eqn{2a2}, taking large radius limit 
involves simultaneously taking $m\to 0$ limit. Note that $m$ is the 
mass parameter in the action. This is the essence of the 
massive 
Penrose limits defined in \eqn{pen1}. Since taking these limits involves
zooming in onto a null geodesic with component along sphere, the 
generalised limits at the basic level 
are the  Penrose limits in \cite{blau,blau1}.} 
This consists in dropping the 
dependence on the coordinates other than $u$. In this way we get
the pp-wave solution  written in Rosen coordinates and 
depending only on $u$, 
\bea
&&\bar g:= dudv + \r~ sin^2 ({u\over2\r}) \sum_{a=1}^{3} 
(dy^a)^2+sin^2({u\ov2})\sum_{a=4}^{8}(dy^a)^2 \br
&&  \bar F_{(4)}= \r~ sin^3({u\ov2\r})~dudy^1dy^2dy^3,~~~\bar\f=0
\eea
with $\bar m=0$ which implies that it is a solution of type IIA instead 
of the massive theory.
We can switch to the new set of (Brinkman) coordinates $(dx^+, 
dx^{-},x^a)$ as in \cite{blau}
\bea
x^{-}=u/2,~~~x^{+}=v -{1\ov4} \sum_a {sin(2\l_a u)\over2\l_a} y^a 
y^a, ~~~
x^a= y^a {sin(\l_au)\over2\l_a} ,
\lll{3a}\eea
we get to the familiar form of the pp-wave metric (a Cahen-Wallach 
spacetime)
\bea\lll{newwave}
&& g:= 2dx^{+}dx^{-} +W(x)  (dx^{-})^2+ \sum_{a=1}^{8}(dx^a)^2\br
&& F_{(4)}= 5dx^{-}dx^1dx^2dx^3
\eea
with $W(x)= -{5\ov2}[(x^1)^2+\cdots+(x^3)^2]-[(x^4)^2+\cdots+(x^8)^2]$.

It can be calculated that the only nonvanishing 
componant of the Ricci 
tensor for the pp-wave metric \eqn{newwave} is 
$$R_{--}={25\ov2}   .$$
As it is  now obvious that the background \eqn{newwave} satisfies 
the field equations 
of ordinary type IIA supergravity and not of the Romans' theory. 
This is because under the Penrose
limits \eqn{pen1}, which effectively takes $m\to 0$, the  massive 
modes have decoupled from the theory.

\subsection{ $16+0$ supersymmetries} 
We now aim to find the amount of supersymmetries preserved by
the pp-wave solution in \eqn{newwave}. 
Let us write down the tangent space metric as 
$$ds^2=2e^{+}e^{-}+ 
e^a~e^a+e^\a~e^\a ,$$
where tangent space indices are taken same as the space-time indices. 
The basis elements are given by 
\be
e^+=dx^{+}+(W/2) 
dx^{-},~e^{-}=dx^{-},~e^a=dx^a .\ee
Only non-vanishing spin connections are $$\o^{+a}={1\ov2}\partial_a W 
dx^{-}.$$
Then the Killing spinors are obtained by solving the vanishing type
IIA supersymmetry 
variations \cite{gia,roma}
\bea
&&\delta \l=0=-{1\ov
  2\sqrt{2}}\partial_\m\f\G^\m\ep-{1\ov2(96)\sqrt{2}} e^{{1\ov4}\f}
F_{\m\n\l\s}\G^{\m\n\l\s} \ep, \label{killing1}   \\
&& \delta \Psi_\m=0= \nabla_\m\epsilon 
+{1\ov256}e^{{1\ov4}\f}
F_{\r\n\l\s}(\G^{\r\n\l\s}_{~~~~~\m}-{20\ov3}\delta^\r_\m\G^{\n\l\s})~\ep
.
\label{killing}\eea
For the pp-wave background \eqn{newwave} the Killing equataion \eqn{killing1} 
implies
\be
\g_{+}\ep=0 \ee
while the Killing equations \eqn{killing} reduce to the following set of equations
\bea
&& \pa_{+}\ep=0,~~~\br
&&[\pa_{-}+{1\ov4}\partial_a W 
~\g_{+}\g^a+{15\ov32}\Theta(\g_{+}\g_{-}-{8\ov3})]\ep=0\br 
&&[\pa_a +{25\ov 32} \g_a\Theta\g_{+}]\ep=0, ~~~{\rm
  for}~a=1,2,3  \br 
&&[\pa_a -{ 15\ov 32} \g_a\Theta\g_{+}]\ep=0,~~~{\rm for}~a=4,...,8 
\eea
where $\Theta=\g_{1}\g_{2}\g_{3}$ and all small $\g$ matrices 
are undressed. Now there is only one type of solutions of the 
above equations. These correspond to
taking $\g_{+}\p=0$ and are called `standard' Killing spinors.\footnote{ We 
are in the frame where $(\g_{+})^2=(\g_{-})^2=0$ and 
$[\g_{+},\g_{-}]_{+}=2$ 
and the projector is $\g_{-}\g_{+}$. } This condition can be satisfied only when
 16 spinors out of the set of total 32 are vanishing.
 For these spinors except for  
$\pa_{-}\ep+\cdots=0$ all other equation are trivially satisfied when we take $\ep$
to be independent of $x^{+}, x^a$. 
These sixteen standard killing spinors are
\be
\ep= e^{{5\ov4}\Theta x^{-}}\p ~ , ~~~~~~~~ \g_{+}\p=0 ~,    
\ee
which depend only on $x^{-}$.
Other set of 
 Killing spinors in the pp-waves are usually those for which $\g_{+}\p\ne 0
 $ and are known as 
`super-numerary'  Killing spinors. 
But these spinors do not exist for this pp-wave solution of type IIA.

\subsection{M-theory pp-waves; Matrix model}
The half-supersymmtric type IIA pp-wave in \eqn{newwave} can be easily 
lifted to  11-dimensional pp-wave solution on a circle which is
\bea\lll{mwave}
&& g:= 2dx^{+}dx^{-} +W(x)  (dx^{-})^2+ \sum_{a=1}^{9}(dx^a)^2\br
&& G_{(4)}= 5dx^{-}dx^1dx^2dx^3\ ,
\eea
with $W(x)= -{5\ov2}[(x^1)^2+\cdots+(x^3)^2]-[(x^4)^2+\cdots+(x^8)^2]$.
This pp-wave in \eqn{mwave} has an overall translational isometry 
direction
$x^9$ along with $SO(3)\times SO(5)$ rotational isometries. The number 
of Killing spinors remains sixteen. Compare
this with maximally supersymmetric 11-dimensional pp-wave solution 
\cite{blau1}
\bea\lll{Hpp}
&& g:= 2dx^{+}dx^{-} +H(x)  (dx^{-})^2+ \sum_{a=1}^{9}(dx^a)^2\br
&&G_{(4)}=5 dx^{-}dx^1dx^2dx^3 ,
\eea
 where the function 
$H= -{25\over9}[(x^1)^2+\cdots+(x^3)^2]-{25\over36}[(x^4)^2+\cdots+(x^9)^2]$. 
This pp-wave has larger rotational isometry group $SO(3)\times SO(6)$. 
The sixteen 
Killing spinors depend upon all the transverse coordinates. While the 
Killing spinors in $1/2$-supersymmetric 
 pp-wave in \eqn{mwave}
do not depend on overall isometry direction $x^9$.
 Thus it appears that getting rid of an overall coordinate, $x^9$, say, 
 in the pp-wave \eqn{Hpp} kills half of 
the supersymmetries.
M-theory pp-wave solutions have been previously studied also 
\cite{hull,clp,singhpp}.

The  BMN matrix model \cite{bmn} on a fully supersymmetric 
pp-wave background admits fuzzy sphere and  5-branes as supersymmetric
solutions. 
Having obtained the 1/2 supersymmetric 11-dimensional pp-wave background 
it would be instructive to discuss the BMN  model  in this 
background. 
For that let us assume $x^{+}\sim 
x^{+} +2\pi R $, being periodic over a circle of radius and 
$R\equiv R_{BFSS}$.    
Then the dynamics of the theory in the momentum sector $p^{-}=p_{+}=N/R$ 
is given by $U(N)$ matrix model action $S=S_{BFSS}+S'$,
\bea
&&S_{BFSS}\sim\int dt~ Tr \bigg[{1\ov2}(D_t x^i)^2 +{1\ov4} [x^i,x^j]^2 
+{\rm fermionic ~terms}\bigg]
\br 
&&S'\sim-\int dt~ {1\ov2}Tr[{5\ov2}(x_1^2+\cdots+x_3^2)+
(x_4^2+\cdots+x_8^2)]+{i\ov3} \ep_{rls}Tr(x^rx^lx^s) \br&&+{\rm 
fermionic~ terms}\bigg] 
\label{mmodel}
\eea
where $i,j=1,2,\cdots,9$; $r,l=1,2,3$ and we have set $R=1,~l_p=1$ and 
time $t$ is identified with $x^{-}$. Comparing it to the BMN matrix 
model, the quadratic mass term for the bosonic  field $x^9$  is 
absent here. The 
isometry group has also been reduced from $SO(6)\times SO(3)$ to 
$SO(5)\times SO(3)$. We conclude that a fuzzy 2-sphere solution still 
exists but with 1/2 supersymmetry. Its relationship with ''giant 
gravitons'' will presumably  be interpretted as delocalised M2-branes 
wrapping the $S^2$. The matrix models of the type 
\eqn{mmodel} with nonconstant fluxes have been
considered in \cite{bonelli}.

\section{Penrose limits in gauged supergravities}
We learned in the previous section that pp-wave backgrounds can be 
obtained by 
applying the generalised plane-wave limits on the $AdS$ vacua of 
 massive IIA supergravity. However, that particular anti-de Sitter 
solution happened to be non-supersymmetric one and the pp-wave had 16
supersymmetry.  In this sense 
there has been an enhancement of supersymmetries, from zero for anti-de 
Sitter space to 
sixteen for the pp-wave, under 
massive Penrose limits. We mantion here the well known 
fact that pp-waves always have 16 or more supersymetries.
The fact that some kind of plane-wave limits exists for massive type-IIA 
vacua 
gives us motivation that these limits should also 
be explored for the gauged supergravities
(supergravities with cosmological constant). Several of the gauged 
supergravity theories
admit Freund-Rubin type {\it electro-vac} and {\it magneto-vac}  
   solutions. 
Again note that the gauged supergravities do have
cosmological constant and therefore cannot 
have pp-wave solutions in them.\footnote{ However, there might exist
 gauged supergravities with potentials (superpotentials) which admit 
 flat vacua in which case they will also have pp-wave solutions in them.}
The idea is the same as before, the application of  generalised Penrose 
limits on the gauged supergravity $AdS$-solutions should give us  
 pp-waves  of 
their ungauged counter parts in an easy and simple manner. Some of these
pp-wave vacua could turn out to be the new solutions. 

It should however be made sure that there exist scaling 
symmetries of the kind we considered in equation \eqn{pen}. 
Practically almost all  gauged 
supergravities are found to have an appropriate scaling symmetry
  under which {\it mass} parameter (cosmological constant)
 gets scaled. 
To emphasize let us note that the gauged  
supergravities, this including the gauged supergravities obtained via 
flux-compactifications, have special couplings of the gauge fields  
introduced through 
covariant derivative $ D \f \sim \pa \f+ g A $ and/or
field strength $F= dA + g [A,A]$. Such couplings have well defined scaling
$A\to \xi^{-1} A,~~~g\to \xi g$ under which $D\f\to D\f$ and $F\to 
\xi^{-1} 
F$ similar to the scalings in \eqn{pen}. 
For these gauged theories  generalised
 Penrose limits  like in  eq. \eqn{pen1} could be defined.
\footnote{If in a gauged supergravity the gauge parameter, $g$, could 
not be scaled, then it would 
not be possible to apply the limits we have discussed in this paper. 
The problem is that even after taking the limits, the cosmological 
constant term in the action (or equations of motion) will servive, 
and there exist no pp-wave solution for the equations of 
motion in presence of cosmological constant.
Though, at the moment, I do not know of an example of such a gauged 
supergravity to offer. } 

The following examples are worth noting in this direction. 
\begin{itemize}
\item
The $6D$ $F(4)$ gauged supergravity \cite{roma2} 
which admits $AdS_2\times S^4$, $AdS_2\times S^2\times S^2$  and 
$AdS_3\times S^3$ will have massive Penrose limits defined. 
There 
exists scaling under which gauge coupling $g$ as well as mass 
parameter will scale to zero value. These vacua under the generalised Penrose limits
will produce pp-wave solutions. 
\item Similarly,
the $5D~N=4$ $SU(2)\times U(1)$ gauged supergravity \cite{roma1}  admits
$AdS_3\times S^2$ and 
 $AdS_2\times S^3$  type electrovac or magnetovac vacua. Some of the
magnetovacs  are supersymmetric with $N\ge1$. There 
exists scaling of the action under which gauge coupling $g$ as well as mass 
parameter scales to zero value. 
The massive 
Penrose limits of these vacua would give  pp-waves of the ungauged
$5D$ supergravity with varying supersymmetries with $N\ge2$. These
pp-waves then can be oxidised to ten dimensions.
\item
Other interesting cases would include the 
case of $4D$ $N=4$ gauged $SU(2)_A \times SU(2)_B$ supergravity  
\cite{fg} which 
admits non-supersymmetric $AdS_2\times S^2$ `Freund-Rubin'
electrovac/magnetovac solutions. 
One can write down Penrose 
limit which will take the gauge couplings $g_A,~g_B$ to zero. These 
Penrose 
limits (accompanied with the scalings of the gauge couplings) will give us 
the pp-wave vacua of the ungauged  $SU(2)\times SU(2)$ theory.\footnote{ 
We note that Freedman-Schwarz model \cite{fg} can be embedded into type-I 
string model compactified on $S^3\times S^3$ \cite{cv}. Under the pp-wave 
limit this internal manifold should become a flat compact six-manifold.}   
\footnote{
In this connection let us note that the Penrose limits of the 
Chamseddine-Volkov monopole 
background \cite{cv} have been  studied in \cite{correa} where 
the authors 
discuss the need of scaling of the cosmological constant. 
 This is quite 
in line with the proposal in the present paper. }    
\end{itemize}

The above examples are only some randomly chosen cases. The gauged 
supergravities in $D=7,~ 8$ and 9 
dimensions also consist interesting examles of anti-de Sitter 
vacua which will lead to pp-waves.

 \section{Conclusions}
In summary, we started with a completely non-supersymmetric $AdS_4\times 
S^6$
solution of massive type-IIA supergravity. After taking appropriate
{\it massive} Penrose limits we have got half-supersymmetric 
pp-wave solution of
the ordinary type IIA supergravity theory.
 
It is  known that taking the Penrose limits \cite{blau1} in no-scale
supergravities leads to the pp-wave vacua. 
The pp-wave solutions always 
have sixteen or more  supersymmetries. 
In this work we have seen that 
the {\it massive} Penrose limits in  massive/gauged 
supergravities 
gets coupled with taking space-time (bulk)  
cosmological constant to zero value. 
In this regard the {\it massive} Penrose limits are the generalisation 
of the pp-wave limits 
of standard (no-scale) supergravity. 
This could also be understood as follows. The
  Penrose limit of anti-de Sitter space amounts to taking large radius 
limit, which 
  effectively means scaling the curvature of spacetime to zero. 
Note that such a limit will then
  simultaneously lead to the flatness of the potential (including 
the potential on the
  moduli space as well). Effectively speaking by taking generalised
pp-wave limit we go from a
  massive case to the massless one, a gauged to ungauged one and from 
flux to the non-flux one. 
This process leads  to the
  recovery of the certain amount (minimum sixteen) of supersymmetries 
into the pp-wave solutions.\footnote{I thank
   Ashoke Sen for an illuminating  discussion on this point.} 
  
It would  be useful to extend this work for a large class of gauged
  supergravities at our disposal. This may provide us with some totally 
new pp-wave solutions unaccounted so far though a large class of 
plane-wave solutions  have
  already been found. It would also be interesting to study generalised 
Penrose limits of gauged supergravities obtained via
flux-compactifications. We intend to report more on this in 
subsequent investigations \cite{follo}.

\leftline{\bf Acknowledgements}
{I gratefully acknowledge the hospitality of  Institute of physics,
  Bhubaneswar as well as Tata Institute of Fundamental 
Research, Mumbai  
where part of this work has been carried out. I would like to thank 
the referee for his careful comments on the draft.}

\vskip .5cm
\appendix{
\section{\underline {\bf Scaling in massive type IIA supergravity }}

The Romans' supergravity theory  \cite{roma} is a
generalization of the type IIA supergravity to include a mass term 
for the NS-NS $ B$-field without 
disturbing the supersymmetry content of the theory. 
The  bosonic action for Romans' theory in the
string frame can be written as (after appropriate rescalings)
\begin{eqnarray} \label{massive2a}
S&=&\int \bigg[ e^{-2\f}\left\{
R~^{\ast}1+4d\f~^{\ast} d\f -
{1\over2} H_{(3)} ~^{\ast} H_{(3)}\right\} 
-{1\over2} F_{(2)} ~^{\ast} F_{(2)} -{1\over2} F_{(4)}
~^{\ast} F_{(4)}- {m^2\ov 2 }~^{\ast} 1 \br
&& +{1\ov2} d C_{(3)} d C_{(3)}  B_{(2)} +{1\ov
2}d C_{(3)} d A_{(1)}  B_{(2)}^2+{1\ov3!}d A_{(1)}
d A_{(1)} B_{(2)}^3
+{1\ov3!}m d C_{(3)} B_{(2)}^3 \br 
&&+{1\ov8}m d A_{(1)}  B_{(2)}^4+ {1\ov40}m^2
B_{(2)}^5\bigg]\ ,
\lll{1aa}
\end{eqnarray}
where $m$ is the mass parameter.\footnote{
Our conventions are same as in
  \cite{hls}  where  every product of forms is understood 
 to be a wedge product. We denote a $p$-form  with a lower index
like $(p)$ which later on is dropped.  }
 The  field strengths in the action \eqn{1aa}
are given by
\bea
 H_{(3)}=d B_{(2)}\ ,
~~~
 F_{(2)}= d A_{(1)} +m B_{(2)}\ ,
~~~
 F_{(4)}= d C_{(3)}
+ B_{(2)}d A_{(1)}+{m\ov2} B_{(2)}^2\ .\br
\lll{a22}\eea
Note that potentials
$ A$ and $ C$  appear only through their derivatives 
in the action \eqn{1aa} and thus obey the standard 
$p$-form gauge invariance $A_{(p)} \to A_{(p)} + d\lambda_{(p-1)}$.
The two-form $  B$ on the other hand also appears
without derivatives but nevertheless the `Stueckelberg'
gauge transformation
\be\label{stueck}
\delta  A=-m\lambda_{(1)}\ ,\qquad
\delta  B=d\lambda_{(1)}\ ,\qquad 
\delta C=-\lambda_{(1)} d A\ 
\ee
leaves the action invariant. This action is presented in a form where one 
can directly implement a massless limit $m \to 0$. The action 
does scale homogeneously as
\be
S\to \xi^{-8} S\ 
\ee
under the scaling 
\bea\label{pena}
 && g_{\m\n}\to \xi^{-2} g_{\m\n}, \qquad \f\to\f\br
&& A_{(1)} \to \xi^{-1} A_{(1)},\qquad B_{(2)} \to \xi^{-2} B_{(2)}, 
\qquad C_{(3)} \to \xi^{-3} C_{(3)} 
\eea
accompanied with the scaling of the mass parameter
\be
 m\to \xi~ m\ .
\ee
The scale parameter  $\xi$ has to be  strictly positive. This 
scaling property of the action
could be used to tune the mass parameter $m$ in the theory
through Penrose limits.

\end{document}